# Metamaterial-based model of the Alcubierre warp drive

Igor I. Smolyaninov

*Department of Electrical and Computer Engineering, University of Maryland, College Park, MD 20742, USA*

**Electromagnetic metamaterials are capable of emulating many exotic space-time geometries, such as black holes, rotating cosmic strings, and the big bang singularity. Here we present a metamaterial-based model of the Alcubierre warp drive, and study its limitations due to available range of material parameters. It appears that the material parameter range introduces strong limitations on the achievable "warp speed", so that ordinary magnetoelectric materials cannot be used. On the other hand, newly developed "perfect" bi-anisotropic non-reciprocal magnetoelectric metamaterials should be capable of emulating the physics of warp drive gradually accelerating up to $1/4 c$.**

Metamaterial optics [1,2] greatly benefited from the field theoretical ideas developed to describe physics in curvilinear space-times [3]. Unprecedented degree of control of the local dielectric permittivity $\varepsilon_{ik}$ and magnetic permeability $\mu_{ik}$ tensors in electromagnetic metamaterials has enabled numerous recent attempts to engineer highly unusual "optical spaces", such as electromagnetic black holes [4-8], wormholes [9], and rotating cosmic strings [10]. Phase transitions in metamaterials are also capable of emulating physical processes which took place during and immediately after the big bang [11,12]. These models can be very informative for phenomena where researchers have no direct experience and therefore limited intuition.

Since its original introduction by Alcubierre [13], the warp drive spacetime has become one of the most studied geometries in general relativity. In the simplest form it can be described by the metric

$$ds^2 = c^2 dt^2 - (dx - v(r)dt)^2 - dy^2 - dz^2 \qquad (1)$$

where $r = \left((x - v_0 t)^2 + y^2 + z^2\right)^{1/2}$ is the distance from the center of the "warp bubble", $v_0$ is the warp drive velocity, and $v = v_0 f(r)$. The function $f(r)$ is a smooth function satisfying $f(0) = 1$ and $f(r) \to 0$ for $r \to \infty$. This metric describes an almost flat spheroidal "warp bubble" which is moving with respect to asymptotically flat external spacetime with an arbitrary speed $v_0$. Such a metric bypasses the speed limitation due to special relativity: while nothing can move with speeds greater than the speed of light with respect to the flat background, spacetime itself has no restriction on the speed with which it can be stretched. One example of fast stretching of the spacetime is given by the inflation theories, which demonstrate that immediately after the big bang our Universe expanded exponentially during an extremely short period of time.

Unfortunately, when the spacetime metric (1) is plugged into the Einstein's equations, it is apparent that exotic matter with negative energy density is required to build the warp drive. In addition, it was demonstrated that the eternal superluminal warp drive becomes unstable when quantum mechanical effects are introduced [14]. Another line of research deals with a situation in which a warp drive would be created at a very low velocity, and gradually accelerated to large speeds. Physics of such a process is also quite interesting [15]. We should point out that the warp drive space-time cannot be reduced to a simple combination of a white-hole and a black-hole event horizons. Such a combination would be non-controversial and "easy" to realize. The difference between the warp drive space-time and such a white-hole/black-hole combination is that the flat space-time region inside the warp bubble is moving as a whole with respect to the flat space-time outside the warp bubble (see metric in eq.(1)) . This non-trivial property of the warp drive space-time has led to conclusion that it cannot be realized even at sub-luminal speeds. Very recently it was demonstrated that even low speed sub-luminal warp drives generically require energy-conditions-violating matter [16]: the $T_{00}$ component of the energy-momentum tensor (the energy density distribution) appears to be negative even at sub-luminal speeds. Therefore, even subluminal warp drives appear to be prohibited by the laws of physics.

In this paper we demonstrate that electromagnetic metamaterials are capable of emulating the warp drive metric (1). Since energy conditions violations do not appear to be a problem in this case, metamaterial realization of the warp drive is possible. Our result is interesting because the body of evidence collected so far seemed to indicate that the warp drives operating at any speed (even sub-luminal) were strictly prohibited by the laws of Nature

Below we will find out what kind of metamaterial geometry is needed to emulate a laboratory model of the warp drive, so that we can build more understanding of the physics involved. It appears that the available range of material parameters introduces strong limitations on the possible "warp speed". Nevertheless, our results demonstrate that physics of a gradually accelerating warp drive can be modeled based on newly developed "perfect" magnetoelectric metamaterials [17]. Since even low velocity physics of warp drives is quite interesting [15,16], such a lab model deserves further study.

To avoid unnecessary mathematical complications, let us consider a 1+1 dimensional warp drive metric of the form

$$ds^2 = (c/n_\infty)^2 dt^2 - (dx - v_0 f(\tilde{x}) dt)^2 - dy^2 - dz^2 \qquad (2)$$

where $\tilde{x} = (x - v_0 t)$ and $n_\infty$ is a scaling constant. In the rest frame of the warp bubble it can be re-written as

$$ds^2 = (c/n_\infty)^2 dt^2 - (d\tilde{x} + v_0 \tilde{f}(\tilde{x}) dt)^2 - dy^2 - dz^2 \qquad (3)$$

where $\tilde{f}(0) = 0$, and $\tilde{f}(\tilde{x}) \to 1$ for $\tilde{x} \to \pm\infty$. The resulting metric is

$$ds^2 = \left( \frac{1}{n_\infty^2} - \frac{v_0^2}{c^2} \tilde{f}^2(\tilde{x}) \right) c^2 dt^2 - d\tilde{x}^2 - 2 v_0 \tilde{f}(\tilde{x}) d\tilde{x} dt - dy^2 - dz^2 \qquad (4)$$

Following ref.[18], Maxwell equations in this gravitational field can be written in the three-dimensional form as

$$\vec{D} = \frac{\vec{E}}{\sqrt{h}} + [\vec{H}\vec{g}], \quad \vec{B} = \frac{\vec{H}}{\sqrt{h}} + [\vec{g}\vec{E}], \qquad (5)$$

where $h = g_{00}$, and $g_\alpha = -g_{0\alpha}/g_{00}$. These equations coincide with the macroscopic Maxwell equations in a magneto-electric material [19]. In the equivalent material

$$\varepsilon = \mu = h^{-1/2} = \frac{1}{\sqrt{\dfrac{1}{n_\infty^2} - \dfrac{v_0^2}{c^2} \tilde{f}^2(\tilde{x})}} \qquad (6)$$

and the only non-zero component of the magneto-electric coupling vector is

$$g_x = \frac{\dfrac{v_0}{c} \tilde{f}(\tilde{x})}{\dfrac{1}{n_\infty^2} - \dfrac{v_0^2}{c^2} \tilde{f}^2(\tilde{x})} \qquad (7)$$

In the subluminal $v_0 << c$ limit eqs.(6) and (7) become

$$\varepsilon = \mu \approx n_\infty \left(1 + \frac{v_0^2 n_\infty^2}{2c^2} \tilde{f}^2(\tilde{x})\right), \quad g_x \approx n_\infty^2 \frac{v_0}{c} \tilde{f}(\tilde{x}) \qquad (8)$$

The magneto-electric coupling coefficients in thermodynamically stable materials are limited by the inequality [20]:

$$g_x^2 \leq (\varepsilon - 1)(\mu - 1) \quad , \qquad (9)$$

which means that a subluminal warp drive model based on the magnetoelectric effect must satisfy inequality

$$\frac{v_0}{c} \tilde{f}(\tilde{x}) \leq \frac{n_\infty - 1}{n_\infty^2} \qquad (10)$$

This inequality demonstrates that while "the true warp drive" in vacuum ($n_\infty = 1$) is prohibited, $n_\infty > 1$ values in a material medium make a warp drive model thermodynamically stable at least at subluminal speeds. This is an important result because the body of evidence collected so far seems to indicate that the warp drives operating at any speed are strictly prohibited by laws of Nature. Equation (10) also provides an upper bound on the largest possible "warp speed", which is achievable within the described metamaterial model. This upper bound is reached at $n_\infty = 2$, and equals to $v_0 = 1/4c$. Therefore, at the very least, we can build a toy model of a warp drive "operating" at $v_0 \sim 1/4c$. Coordinate dependence of the metamaterial parameters in such a model is shown in Fig.1 assuming $\tilde{f}(\tilde{x}) = \left(1 + a^2 / \tilde{x}^2\right)^{-1}$.

However, in classical magnetoelectric materials, such as $Cr_2O_3$ and multiferroics, actual values of magnetoelectric susceptibilities are two orders of magnitude smaller than the limiting value described by eq. (9) [21], so that the warp drive model is impossible to make with ordinary materials. On the other hand, recently developed "perfect" magnetoelectric metamaterials [17], which can be built based on such designs as split ring resonators, fishnet structures [22], etc. allow experimentalists to reach the limiting values described by eq.(9), and make a lab model of the warp drive possible. Following ref.[17], the effective susceptibilities of the split ring metamaterial can be written in the RLC-circuit model as

$$\varepsilon = 1 + \frac{nCd^2\omega_0^2}{\left(\omega_0^2 - \omega^2 - i\omega\gamma\right)}, \quad \mu = 1 + \frac{nCS^2\omega^2\omega_0^2}{c^2\left(\omega_0^2 - \omega^2 - i\omega\gamma\right)}, \text{ and} \quad (11)$$

$$g = \frac{inCdS\omega\omega_0^2}{c\left(\omega_0^2 - \omega^2 - i\omega\gamma\right)}, \quad (12)$$

where $n$ is the split ring density, $d$ is the gap in the ring, $S$ is the ring area, and $C$ is the gap capacitance. These expressions explicitly demonstrate that the split ring metamaterial considered in ref. [17] satisfies the upper bound given by eq.(9), and therefore can be used as one of the building blocks in the metamaterial warp drive design. On the other hand, this particular split-ring design cannot be used without modification, since this metamaterial is reciprocal.

An actual laboratory demonstration of a metamaterial warp drive space time would require a non-reciprocal bi-anisotropic metamaterial, in which both spatial and time reversal symmetries are broken. In addition, the metamaterial loss issue has to be overcome. Since the issue of metamaterial loss compensation using gain media is well studied (see for example recent experimental demonstration of loss compensation in a negative index metamaterial [23]) let us concentrate on the experimental ways of breaking spatial and time reversal symmetries. Breaking the mirror $x \leftrightarrow$ -$x$ symmetry is most easily achieved by deformation of the metamaterial, which can be easily done in either one of the most popular split ring [17] or fishnet [22] metamaterial designs. As far as breaking the time-reversal $t \leftrightarrow$ -$t$ symmetry is concerned, there are two most natural ways to break this symmetry in solids: application of external magnetic field [24,25], or spin-orbit interaction in a nonsymmorfic lattice [26]. In addition, such chiral superconductors as $Sr_2RuO_4$ [27] may be used in superconducting metamaterial designs [28]. Application of external electric and magnetic fields is known to break both spatial and time symmetries of such materials as methyl-cyclopentadienyl-Mn-tricarbonil molecular liquids [24], thus creating an illusion of a moving (non-reciprocal bi-anisotropic) medium [25]. Experimental results of ref. [24] do indeed demonstrate this behavior (however, the emulated "medium velocity" is very low, of the order of 50 nm/s [25]). Since utilization of magnetized particles, such as ferrites is easily applicable in the metamaterial design, all the ingredients necessary for experimental realization of the Alcubierre metric have been demonstrated in the experiment. Moreover, we should point out that very recently it was also asserted [29,30] that material parameters, which are necessary to achieve a warp drive imitation in a nanostructured metamaterial are indeed possible.

Let us come back to the split ring-based "perfect magnetoelectric metamaterial" design implemented in [17], and demonstrate how the time-reversal symmetry may be broken in this metamaterial. One of the possible metamaterial geometries is shown schematically in Fig.2. An elementary unit of the split ring-based "perfect magnetoelectric metamaterial" design of ref.[17] is supplemented with a magnetized ferrite particle. The particle is magnetized and shifted along the $x$-direction with respect to the center of the split ring. Thus, this geometry explicitly violates spatial and time reversal symmetries, resulting in a non-reciprocal bi-anisotropic metamaterial. As demonstrated in [31], near the ferromagnetic resonance frequency in such a metamaterial

$$g \sim \frac{\omega_m \omega_0}{\omega_0^2 - \omega^2},\quad (13)$$

where $\omega_0$ is the ferromagnetic resonance frequency, and $\omega_m = \gamma M_0$. Thus, in the design presented in Fig.2 $g_x$ is proportional to the particle magnetization $M_0$ in a given location, which explicitly demonstrate the non-reciprocal nature of this metamaterial design. Time reversal $t \leftrightarrow -t$ leads to change of sign of $g_x$. We should also note that very recently a somewhat related metamaterial design has been proposed in ref.[32] , which emulates medium motion at an arbitrary speed. While demonstrating the proof of principle, the designs presented in Fig.2 and ref.[32] may only be considered as a first step. Such complicated metamaterial designs typically contain many unwanted terms in $\varepsilon$, $\mu$ and $g$, which must be carefully eliminated by iteration, so that the ideal form of eqs.(8) may be achieved.

Light rays propagation inside the metamaterial model of the warp drive "operating" at $1/4c$ is illustrated in Fig.3. The metamaterial medium (1) outside the warp bubble is engineered to have properties of a medium moving towards the warp bubble with the designed warp speed, while medium (3) is "moving" away from the bubble. In the reference frame moving with the warp speed these media look exactly the same as medium (2) at rest. Ray trajectories were calculated assuming a step-like $\tilde{f}(\tilde{x})$ profile. Rays are emitted by a point source located at the origin point (0,0,0) of the coordinate frame inside the "warp bubble" (marked as medium (2)). Boundaries of the warp bubble are located at x=±5 (they are marked by the dashed lines). At large enough incidence angles light rays originating inside the "warp bubble" cannot penetrate into medium (3) (in the hypothetical superluminal warp drive this would be true for any incidence angle: this boundary would look like a white hole event horizon). On the other hand, all the light rays propagating towards the other boundary of the "warp bubble" can propagate into medium (1). Note that the metamaterial medium (1) is identical to medium (3).

In conclusion, we have presented a metamaterial-based model of the Alcubierre warp drive metric. It appears that the material parameter range introduces strong limitations on the achievable "warp speed", so that ordinary magnetoelectric materials cannot emulate the warp drive. On the other hand, newly developed "perfect" magnetoelectric bi-anisotropic non-reciprocal metamaterials should be capable of emulating the physics of gradually accelerating warp drive, which can reach "warp speeds" up to $1/4c$.

**Figure Captions**

**Figure 1.** Spatial distributions of $\varepsilon$, $\mu$, and $g_x$ in the metamaterial model of a warp drive gradually accelerated up to $1/4c$.

**Figure 2.** Example of a non-reciprocal bi-anisotropic metamaterial geometry, which explicitly violates spatial and time reversal symmetries. An elementary unit of the split ring-based "perfect magnetoelectric metamaterial" design of ref.[17] is supplemented with a magnetized ferrite particle. The particle is magnetized and shifted along the $x$-direction with respect to the center of the split ring. The particle magnetization is proportional to the required $g_x$ in a given location.

**Figure 3.** Light rays propagation inside the metamaterial model of the warp drive operating at $1/4c$. Rays are emitted by a point source located at the origin point (0,0,0) of the coordinate frame inside the "warp bubble". Boundaries of the warp bubble are located at x=±5. Note that the metamaterial media (1) and (3) are identical.

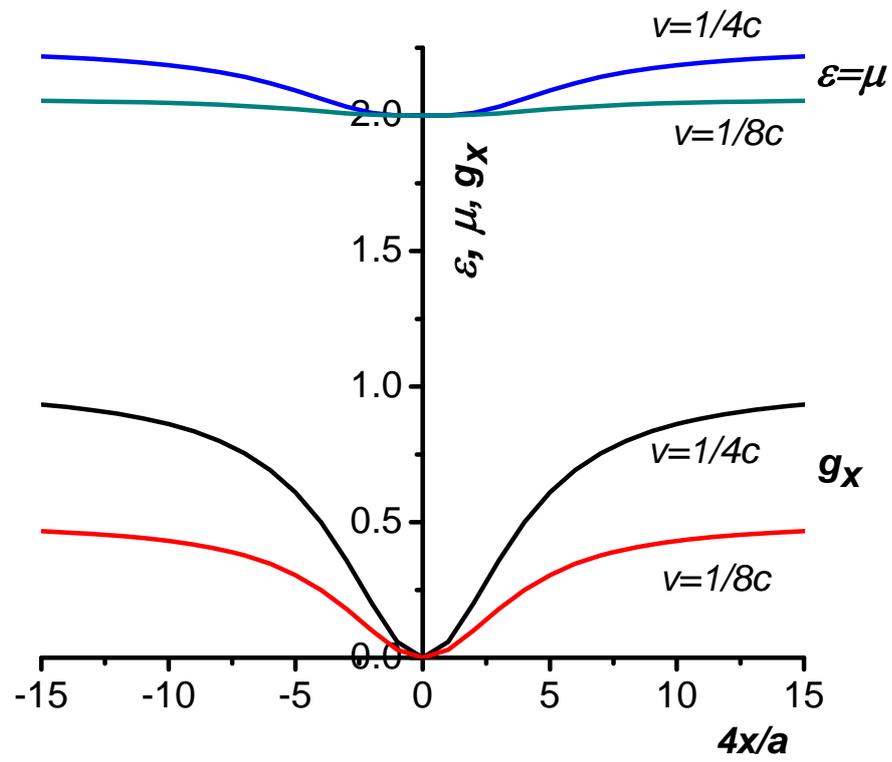

Fig.1

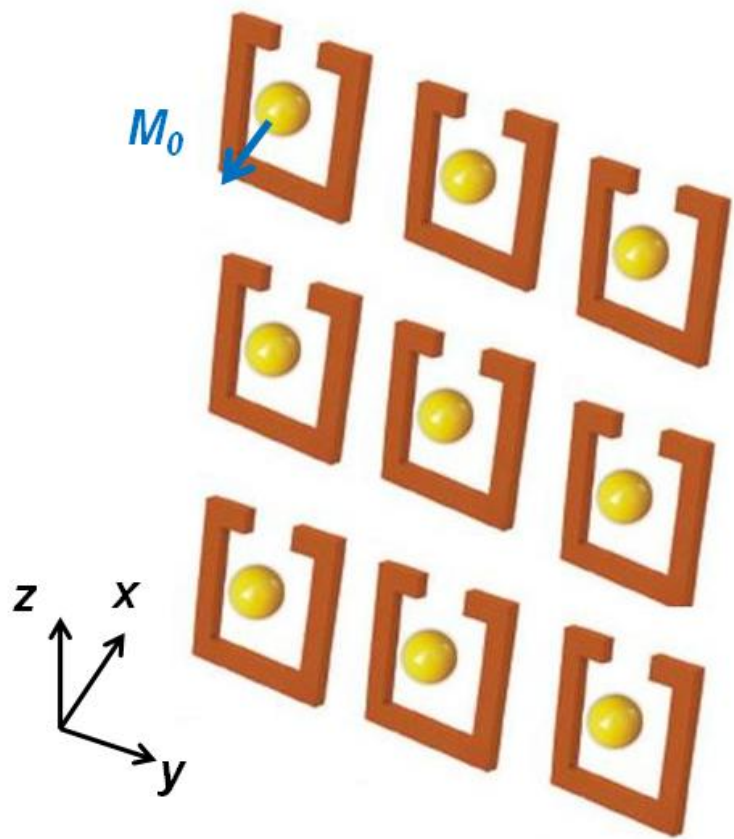

Fig.2

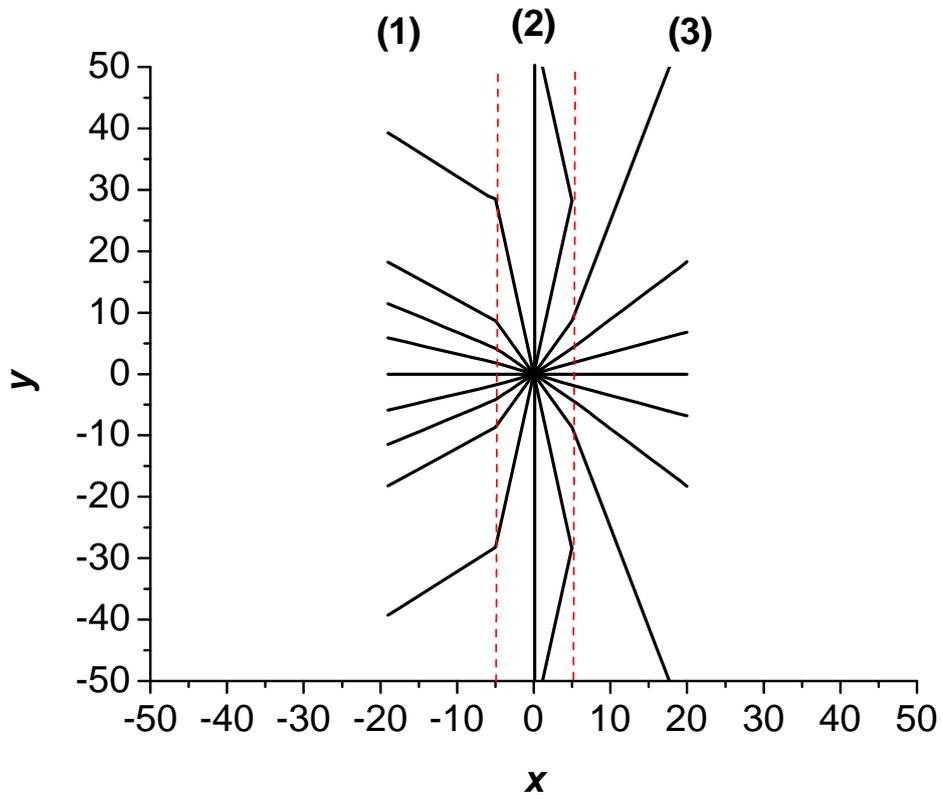

Fig.3